\documentclass[epj]{webofc}
\usepackage[utf8]{inputenc}
\usepackage[varg]{txfonts}   
\usepackage{booktabs}
\usepackage{xcolor}
\usepackage{tikz}
\usetikzlibrary{decorations.pathreplacing}

\definecolor{darkred}{rgb}{0.4,0.0,0.0}
\definecolor{darkgreen}{rgb}{0.0,0.4,0.0}
\definecolor{darkblue}{rgb}{0.0,0.0,0.4}
\usepackage[bookmarks,linktocpage,colorlinks,
    linkcolor = darkred,
    urlcolor  = darkblue,
    citecolor = darkgreen]{hyperref}

\newcommand{\beq}{\begin{eqnarray}}
\newcommand{\eeq}{\end{eqnarray}}
\newcommand{\tr}{\ensuremath{\mathrm{Tr}}}
\newcommand{\ie}{{\it{i.e.} }}
\newcommand{\wrt}{{\it{w.r.t.} }}

\usepackage{subfigure}

\wocname{EPJ Web of Conferences}
\woctitle{Lattice2017}

\begin{document}

\selectlanguage{english}

\title{
Flux tubes in $N_f=2+1$ QCD with external magnetic fields
}

\author{
\firstname{Claudio} \lastname{Bonati}\inst{1} \fnsep\thanks{Present address: 4,5} \and
\firstname{Salvatore} \lastname{Cal\`i}\inst{2,3} \and
\firstname{Massimo} \lastname{D'Elia}\inst{4,5} \and
\firstname{Michele} \lastname{Mesiti}\inst{4,5} \fnsep\thanks{Present address: Swansea University, Singleton Park, Swansea SA2 8PP, Wales (UK).} \and
\firstname{Francesco}  \lastname{Negro}\inst{4}\fnsep\thanks{Speaker, \email{fnegro@pi.infn.it}} \and
\firstname{Andrea} \lastname{Rucci}\inst{4,5}\and
\firstname{Francesco} \lastname{Sanfilippo}\inst{6}
}

\institute{
Dipartimento di Fisica e Astronomia and INFN (Firenze), Via G. Sansone 1, 50019, Sesto Fiorentino (Italy)
\and
University of Cyprus, P.O. Box 20537, 1678 Nicosia, (Cyprus)
\and
University of Wuppertal, Gau\ss str. 20, 42119 Wuppertal, (Germany)
\and
INFN - Sezione di Pisa, Largo B. Pontecorvo 3, 56127 Pisa (Italy)
\and
Dipartimento di Fisica dell'Universit\`a di Pisa, Largo B. Pontecorvo 3, 56127 Pisa (Italy)
\and
INFN - Sezione di Roma Tre, Via della Vasca Navale 84, 00146 Roma (Italy)
}

\abstract{
We study the behavior of the confining flux tube in 
$N_f=2+1$ QCD at the physical point, discretized with
the stout smearing improved staggered quark action
and the tree level Symanzik gauge action.
We discuss how it depends on a uniform external
magnetic field, showing how it displays anisotropies
with respect to the magnetic field direction.
Moreover, we compare the observed anisotropy pattern
with that of the static quark-antiquark ($Q\overline{Q}$)
potential we obtained in~\cite{Bonati:2014ksa,Bonati:2016kxj}.
}

\maketitle

\section{Introduction and numerical setup}\label{sec-1}
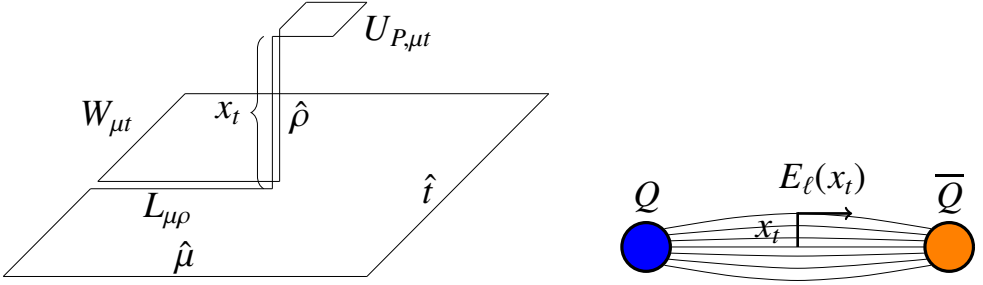
\begin{figure}[h] 
  \centering
\scalebox{0.8}{
\begin{tikzpicture}
\draw[]  ( 0.00, 0.00)--( 6.00, 0.00);
\draw[]  ( 6.00, 0.00)--( 9.00, 3.00);
\draw[]  ( 9.00, 3.00)--( 3.00, 3.00);
\draw[]  ( 3.00, 3.00)--( 1.56, 1.56);
\draw[]  ( 1.56, 1.56)--( 4.56, 1.56);
\draw[]  ( 4.56, 1.56)--( 4.56, 4.06);
\draw[]  ( 4.56, 4.06)--( 5.00, 4.50);
\draw[]  ( 5.00, 4.50)--( 6.00, 4.50);
\draw[]  ( 6.00, 4.50)--( 5.44, 3.94);
\draw[]  ( 5.44, 3.94)--( 4.44, 3.94);
\draw[]  ( 4.44, 3.94)--( 4.44, 1.44);
\draw[]  ( 4.44, 1.44)--( 1.44, 1.44);
\draw[]  ( 1.44, 1.44)--( 0.00, 0.00);
\draw [decorate,decoration={brace,amplitude=6pt},xshift=-4pt,yshift=0pt] ( 4.44, 1.44) -- ( 4.44, 3.94) node [black,midway,xshift=-0.6cm] {\LARGE $x_t$};
\draw ( 3.00, 0.70) node[anchor=north]{{\LARGE $\hat \mu$}};
\draw ( 7.00, 1.80) node[anchor=north]{{\LARGE $\hat t$}};
\draw ( 4.90, 3.00) node[anchor=north]{{\LARGE $\hat \rho$}};
\draw ( 2.70, 1.50) node[anchor=north]{{\LARGE $L_{\mu\rho}$}};
\draw ( 1.70, 3.00) node[anchor=north]{{\LARGE $W_{\mu t}$}};
\draw ( 6.50, 4.40) node[anchor=north]{{\LARGE $U_{P,\mu t}$}};
\end{tikzpicture}
\hspace{1.0cm}
\begin{tikzpicture}
\coordinate (center1) at (0,2);
\coordinate (center2) at (5,2);
\fill[blue] (center1) arc (0:360:0.4);
\fill[orange]  (center2) arc (0:360:0.4);
\draw[black,line width=1.5pt] (center1) arc (0:360:0.4);
\draw[black,line width=1.5pt] (center2) arc (0:360:0.4);
\draw ( -0.40, 3.20) node[anchor=north]{{\LARGE $Q$}};
\draw ( 4.60, 3.35) node[anchor=north]{{\LARGE $\overline{Q}$}};
\draw ( 2.50, 3.50) node[anchor=north]{{\LARGE $E_{\ell}(x_t)$}};
\draw[->,line width=1.25pt]  ( 2.10, 2.55)--( 3.00, 2.55);
\draw ( 1.6, 2.55) node[anchor=north]{{\LARGE $x_t$}};
\draw[-,line width=1.25pt]  ( 2.10, 2.00)--( 2.10, 2.57);
\draw [black] plot [smooth] coordinates {
( 0.000 , 2.000)
( 4.200 , 2.000)
 };
\draw [black] plot [smooth] coordinates {
(-0.02,2.10)
( 2.10,2.15)
(4.22,2.10)
 };
\draw [black] plot [smooth] coordinates {
(-0.02,1.90)
( 2.10,1.85)
(4.22,1.90)
 };
\draw [black] plot [smooth] coordinates {
(-0.03,2.20)
( 2.10,2.35)
(4.23,2.20)
 };
\draw [black] plot [smooth] coordinates {
(-0.03,1.80)
( 2.10,1.65)
(4.23,1.80)
 };
\draw [black] plot [smooth] coordinates {
(-0.1,2.30)
( 1.05,2.48)
( 2.10,2.55)
( 3.15,2.48)
(4.3,2.30)
 };
\draw [black] plot [smooth] coordinates {
(-0.1,1.70)
( 1.05,1.52)
( 2.10,1.45)
( 3.15,1.52)
(4.3,1.70)
 };
\end{tikzpicture}
}
  \caption{Schematic representation of the observable in Eq.~(\ref{connanddisc}) that we adopted
to determine the color-electric field $E_\ell(x_t)$ between two static color sources on the lattice. }
  \label{correlator}
\end{figure}

The study of flux tubes forming between pairs of static color sources 
have always been a way to study the emergence of a confining potential in QCD.
This is rather independent of the particular confining mechanism, even if the very idea 
of flux tube formation emerges very naturally within the dual
superconductor scenario for color confinement.\\
The purpose of this paper is to provide a first investigation
of flux tube formation in the presence of a magnetic background field.
Various studies~\cite{Miransky:2002rp,Chernodub:2010bi,Galilo:2011nh,Ozaki:2013sfa,Bonati:2014ksa,Bonati:2016kxj}
have shown that the confining potential gets strongly modified
by the magnetic field, and that the string tension
parallel and perpendicular to the magnetic field differ;
moreover, for large enough magnetic fields
the former could even vanish~\cite{Endrodi:2015oba,Bonati:2016kxj}.
Within this context, looking at the flux tube 
provides a new way to achieve a better comprehension of how the 
magnetic field acts on the confining properties
of QCD.\\
The color-electric field $E_{\ell}(x_t)$ which develops between a pair 
of static color sources can be evaluated on the lattice
by considering the connected correlator sketched in Fig.~\ref{correlator},
that has been originally introduced in Ref.~\cite{DiGiacomo:1989yp,DiGiacomo:1990hc} and subsequently
adopted in various works~\cite{Cea:2017ocq,Cea:2012qw}.
The subscript $\ell$ stands for {\it longitudinal}, since we are considering
the field component longitudinal with respect to the interquark separation.
It has been observed that the transverse components of the color-electric field and all the color-magnetic
field components are negligible within the flux tube~\cite{Cea:1995zt}.
We compute the field exactly at the mid-point of the $Q\overline{Q}$ separation
and at various {\it transverse} distances $x_t$. In such a way, we can reconstruct
the profile (or shape) of the flux tube.\\
$E_{\ell}(x_t)$ can be expressed in terms of the
connected correlator $C_{\mu t}(x_t\hat\rho)$ and of its disconnected counterpart $D_{\mu t}(x_t\hat\rho)$:
\beq
\label{connanddisc}
\centering
C_{\mu t}(x_t \hat \rho) = \langle \tr (W_{\mu t} L_{\mu \rho} U_{P,\mu t} L_{\mu\rho}^\dagger ) \rangle
\ \ \ \ \ \ {\rm and} \ \ \ \ \ \ D_{\mu t}(x_t \hat \rho) = \langle \tr (W_{\mu t}) \tr(U_{P,\mu t}) \rangle,
\eeq
which consist of a gauge invariant product of link variables
conveniently splittable into $3$ constituent paths: $W_{\mu t}$,
$L_{\mu\rho}$, and $U_{P,\mu t}$.
$W_{\mu t}$ is a Wilson loop in the $\mu t$ plane ($\mu$ is one spatial
direction while $t$ is the temporal direction) with temporal extent
$T = n_T a$ and spatial extent $R = n_R a $ along $\mu = X$, $Y$ or $Z$;
it represents two static color sources separated by a spatial distance
$R$ along $\mu$, propagating for a euclidean time $T$.
The plaquette $U_{P,\mu t}$ is located in the position where the color-electric
field is probed: inbetween the sources (\ie in the middle of the Wilson loop)
and shifted by $x_t$ along the {\it transverse} spatial direction $\hat \rho\neq \hat \mu$.
Finally, $L_{\mu\rho}$ is an L-shaped Schwinger line starting from the middle of
one of the temporal sides of $W_{\mu t}$, reaching the center of the Wilson Loop and
ending in the point where $U_{P,\mu t}$ insists on. The explicit expression of the color-electric field observable in the $Q\overline{Q}$ background on the lattice reads
\beq
\label{Fmunu}
F_{\mu t}(x_t \hat \rho) =   \frac{1}{a^2g} \left(\frac{C_{\mu t}(x_t \hat \rho) }{\langle \tr(W_{\mu t})\rangle } -  \frac{1}{N_c} \frac{D_{\mu t}(x_t \hat \rho) }{\langle \tr(W_{\mu t})\rangle } \right),
\eeq
where $a$ is the lattice spacing and $g=\sqrt{6/\beta}$ is the gauge coupling.
Eq.~(\ref{Fmunu}) gives directly a determination of the chromoelectric field itself,
while alternative definitions based on the disconnected correlator $D_{\mu\rho}(x_t) $
allows only to determine the expectation value of the square of the field~\cite{Fukugita:1983du,Cardoso:2013lla}.\\
If we consider the case without the external field, \ie the case at $eB=0$, all
the interquark separation directions are equivalent, as well as all the possible
transverse directions. Hence, we can perform an average over all the $6$ possible
combinations of directions and define $E_{\ell}(x_t)$ as
\beq
E_{\ell}(x_t) \equiv \frac{1}{6} \left( F_{xt}(x_t \hat y) + F_{xt}(x_t \hat z) + F_{yt}(x_t \hat x) + F_{yt}(x_t \hat z) + F_{zt}(x_t \hat x) + F_{zt}(x_t \hat y) \right).
\eeq
Conversely, when we have a nonzero external field $eB\neq 0$ oriented along the $\hat z$ axis
we cannot perform such an average because the lattice rotational symmetry is explicitly broken.
Then, we identify 3 different classes of correlators which can be safely averaged, according
to their parallel ($\parallel$) or transverse ($\perp$) orientation with respect to the $B$ field.
These classes are reported in the following table, where the ``representative of the class'' entries
contain the symbol that we will use to identify the data of each class.
\begin{center}
\begin{tabular}{|c|c|c|c|}\hline
  $\vec{B}$ and $\hat \mu$  & $\vec{B}$ and $\hat\rho$  & Representative of the class & Definition of the class  \\\hline
  $\perp$     & $\perp$     & $XT - Y$ & $E_{x}(x_t\hat y) \equiv  \frac{1}{2}\left( F_{xt}(x_t \hat y) + F_{yt}(x_t \hat x)  \right)$ \\
  $\perp$     & $\parallel$ & $XT - Z$ & $E_{x}(x_t\hat z) \equiv  \frac{1}{2}\left( F_{xt}(x_t \hat z) + F_{yt}(x_t \hat z)  \right)$ \\
  $\parallel$ & $\perp$     & $ZT - X$ & $E_{z}(x_t\hat x) \equiv  \frac{1}{2}\left( F_{zt}(x_t \hat x) + F_{zt}(x_t \hat y)  \right)$ \\\hline
\end{tabular}
\end{center}
\subsection{Smearing}\label{sec-1-1}
\vspace{-0.2cm}
The observable we are interested in consists of the
product of several link variables along the path described in the previous
section and shown in Fig.~\ref{correlator}. Hence, as well as for the
determination of the Wilson Loop, it is convenient to perform some
sort of smearing on the SU(3) link variables,
in order to reduce the UV-fluctuations and gain a better
signal-to-noise ratio.
To this aim, following previous works~\cite{Cea:2017ocq}, we adopted a smearing
procedure with one single HYP smearing step for the temporal links~\cite{Hasenfratz:2001hp}
with coefficients $\vec{\alpha} = (1,0.5,0.5)$ and $N_{APE}$ APE smearing steps for the spatial links~\cite{Albanese:1987ds}
\beq
U_\mu^{APE} = {\rm Proj}_{SU(3)}\left( U_\mu + \alpha_{APE} V_\mu \right),
\eeq
with $\alpha_{APE} = 1/6$ and where $V_\mu$ is the sum of all the four spatial $3$-links staples related to $U_\mu$.\\
We evaluated our observables for several different values
of $N_{APE}$, in order to be able to address the issue of the dependence
of the results on the smearing level.\\
The APE smearing procedure is empirically equivalent~\cite{Alexandrou:2017hqw}
to other smoothing techniques, like cooling, Wilson flow,
stout smearing, and other kinds of smearing, which all can be
interpreted as diffusive processes. Hence, it is convenient
to translate the number of APE steps in a smearing radius in physical
units which can be compared to other smearing procedures.
Following~\cite{Alexandrou:2017hqw}, we adopted the formula
\beq
R_s = a \sqrt{\frac{8\  \alpha_{APE}\ N_{APE}} {1+6\ \alpha_{APE}} }
\eeq
which holds in the case of a four-dimensional smearing, while here we are
performing 3D smearing. Even if it is possible to derive a relation
also for the 3D case, here we are just interested in giving the
order of magnitude of $R_s$, which should be relatively similar with respect to the 4D case.
\subsection{Lattice discretization}\label{sec-1-2}
\vspace{-0.2cm}
We discretized the $N_f = 2+1$ QCD lagrangian density at the physical point 
adopting, as in~\cite{Bonati:2014ksa,Bonati:2016kxj}, improvements both in the gauge and in the fermion sectors.
Moreover, we took into account the effect of the background constant and uniform (electro-)magnetic field,
which is directly coupled to the fermionic degrees of freedom.
This can be attained by adding the abelian gauge potential term
$iq_fA_\mu$ to the gauge covariant derivative, so that its complete expression
reads $D_\mu = \partial_\mu+igA_\mu^aT^a+iq_fA_\mu$,
where $q_f$ is the electric charge of a given quark flavour and $A_\mu^a$ is the non-Abelian gauge potential.
In the discrete formalism of Lattice QCD, such a modification of $D_\mu$  can be introduced by
multiplying the usual SU(3) variables that appear in the Dirac operator by proper $U(1)$ phases $u_{i;\mu}^f = \exp(i q_f a A_\mu(i))$, where $A_\mu$ is a four-potential for a uniform magnetic field along the $Z$ axis. All the phases that differs from the identity are:
\beq
u_{i;y}^f = e^{i a^2 q_f B_z i_x}\ \ \ {\rm and}\ \ \ \left.u_{i;x}^f\right|_{ix=Lx} = e^{-i a^2 q_f L_x B_z i_y} .
\eeq
This choice describes a uniform magnetic field only if $B_z$ satisfies the quantization condition~\cite{AlHashimi:2008hr}: $e B = 6\pi b_z/(a^2 N_x N_y)$. Within this setup, the euclidean partition function reads:
\begin{eqnarray}
\label{partfunc}
\hspace{-0.7cm}\mathcal{Z}(B) \!\!\!\!\!&=&\!\!\!\! \int \!\mathcal{D}U \,e^{-\mathcal{S}_{Y\!M}} \!\!\!\!\prod_{f=u,\,d,\,s} \!\!\! \det{({D^{f}_{\textnormal{st}}[B]})^{1/4}},
\end{eqnarray}
where by $\mathcal{D}U$ we mean the functional integration over the SU(3) links.
$\mathcal{S}_{Y\!M}$ is the tree level improved Symanzik gauge action~\cite{Weisz:1982zw,Curci:1983an}.
The nonabelian gauge variables
in the staggered Dirac operator, $D^f_{\textnormal{st}}$, are the two times stout-smeared links~\cite{Morningstar:2003gk}.
See~\cite{Bonati:2014ksa,Bonati:2016kxj} for further details.
\vspace{-0.2cm}
\section{Numerical results}\label{sec-2}
\vspace{-0.2cm}
We consider the same setup we adopted in~\cite{Bonati:2014ksa,Bonati:2016kxj},
where we simulated the discretization of QCD in Eq.~(\ref{partfunc}) with the RHMC algorithm on several lattices, at several lattice spacings and in correspondence of different values of the magnetic field.
Here we focus on the case of the simulations performed on a $48^3\times 96$ lattice
at $\beta = 3.85$ and with bare quark masses
chosen in order to be at the physical point.
The corresponding lattice spacing in physical units is $a \simeq 0.0989$ fm, while
the four-volume is about $(5\ {\rm fm})^3 \cdot 10 \ {\rm fm}$.
In order to address the effect of the external field on the flux tube profile,
we considered the case  $eB=0$ and also 5 non-vanishing values of $eB$, up to $\sim\ 3 {\rm GeV}^2$.
For each field we considered $\mathcal{O}(30)$ thermalized configurations,
separated by 25 steps of RHMC algorithm, and computed the flux tube every 10
steps of APE smearing, with $N_{APE}$ spanning from 10 up to 250.
\subsection{Numerical results at $eB = 0$}\label{sec-2-1}
\begin{figure}[b] 
  \centering
 \includegraphics[width=0.45\textwidth]{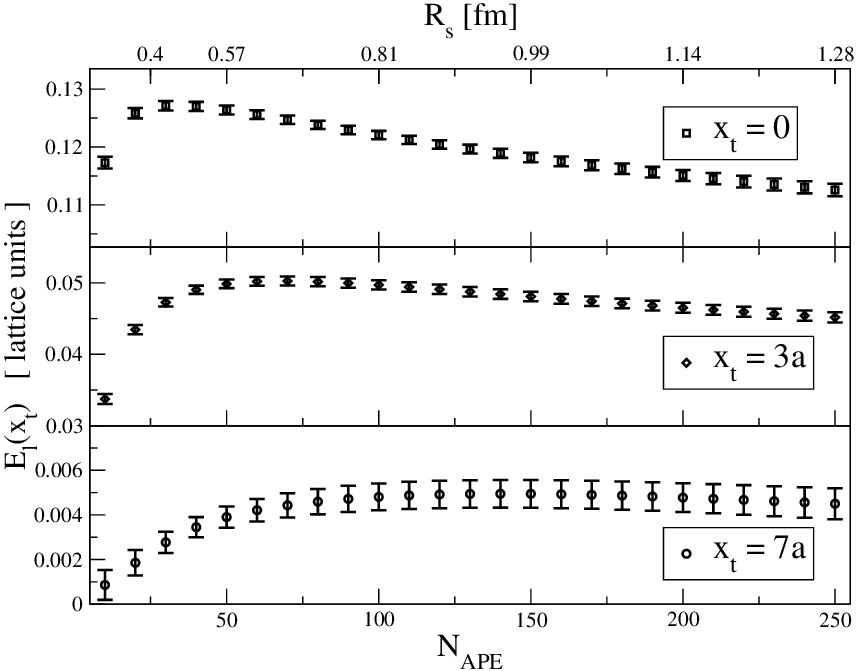}
  \caption{Dependence of the color-electric field strength on $N_{APE}$ (and on $R_s$). Data refers to a $7\times 7$ Wilson
Loop, \ie to an interquark separation of about $0.7$ fm.}
  \label{tube_B0_vs_nape_vs_Rs}
\end{figure}

\vspace{-0.2cm}
In Fig.~\ref{tube_B0_vs_nape_vs_Rs}, we plot the longitudinal color-electric
field $E_\ell(x_t)$ as a function of the smearing level for a $Q\overline{Q}$ separation
of about $0.7$ fm (\ie with a $7\times 7 $ Wilson Loop), for three values of transverse distance $x_t = 0,\ 3a,\ 7a$.
Although it would be very interesting to study how the flux tube depends
on the distance between the color sources (\ie on the size of the
Wilson Loop), we report here preliminary results for this single interquark separation,
and we leave the study of such a dependece to future works.
Data show a significant dependence on $N_{APE}$, \ie the smearing radius $R_s$:
the color-electric field strength initially grows, then it reachs a sort of maximum/plateaux, but soon after
it starts again to decrease. Such a behaviour is similar to that observed for
the field strenght correlators~\cite{DElia:1997sdk,DElia:2015eey}.
Moreover, the larger the value of $x_t$ the larger the smearing radius $R_s$ where we observe the maximum/plateaux.
Hence, we need to choose a specific prescription to fix the value of $R_s$.
In Ref.~\cite{Cea:2017ocq}, the prescription is to take the value at the maximum/plateaux;
this is analogous to what has been done in the literature for similar quantities like
the gauge-invariant field strength correlators~\cite{DElia:1997sdk,DElia:2015eey}.
Anyhow, this prescription implies that the field value is taken at a
different number of smearing steps according to the value of $x_t$.
\begin{figure}[t] 
  \centering
 \includegraphics[width=0.35\textwidth]{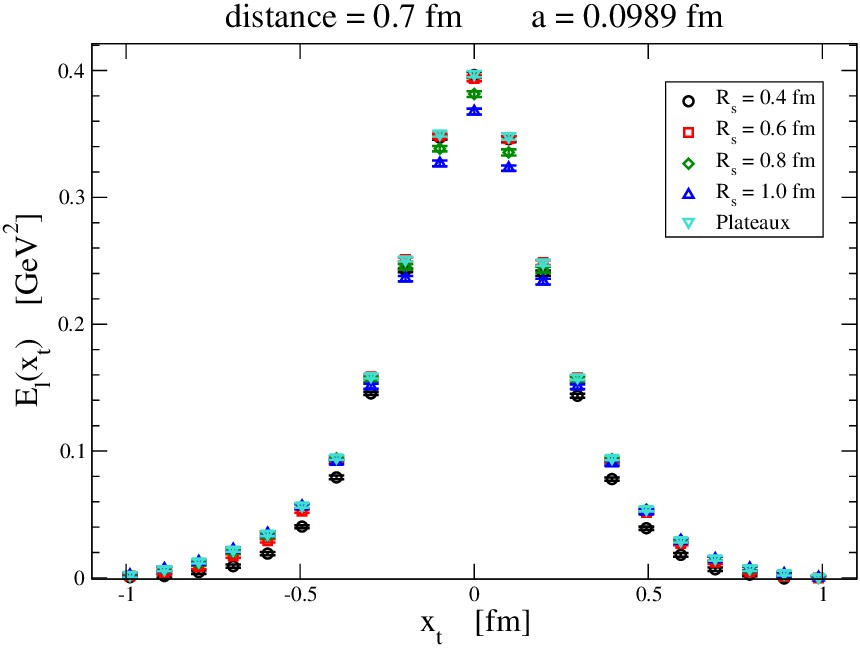}
 \includegraphics[width=0.35\textwidth]{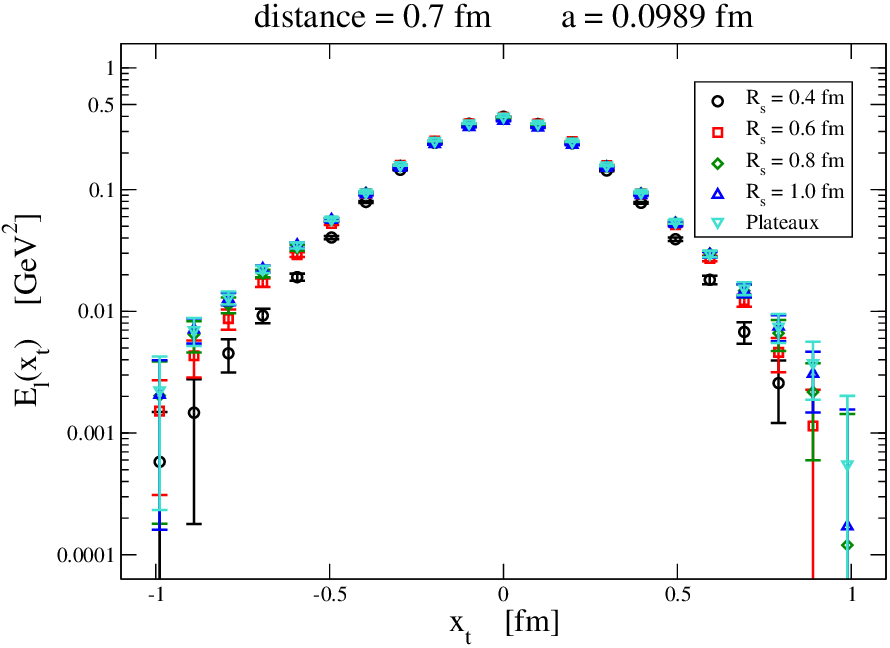}
  \caption{The shape of the flux tube at $eB=0$ generated by two static
color sources separated by $0.7$ fm as obtained with the fixed $R_s$ method and with the maximum/plateaux method
plotted in linear scale ({\it Left}) and in log scale ({\it Right}).}
\vspace{-0.5cm}
  \label{compare_tube_B0}
\end{figure}
In order to check for possible systematics, it is important to consider also  different prescriptions,
to consistently perform the continuum limit and, finally, to compare them.
Among the possible choices, here we consider the case in which
all the field strength values are measured at the same fixed value of the smearing
level $R_s$, in physical units, for all the transverse separations $x_t$.
Then, the continuum limit should be taken at fixed $R_s$, and continuum
results obtained for different values of $R_s$ should be compared among them and with the plateaux method.
Such a prescription is similar to what is done for other correlators 
using the gradient flow as a smoothing technique, and is 
actually the same thing in view of the equivalence between 
all smoothing techniques~\cite{Bonati:2014tqa,Alexandrou:2017hqw}.
In Fig.~\ref{compare_tube_B0}, we compare the flux tube profile for
different prescriptions for fixing $R_s$, including the maximum/plateaux method.
Since the comparison is limited to one lattice spacing and no continuum
limit is performed, we cannot conclude anything about the fate of the observed discrepancies.
We leave such a study to a future work and
we limit ourselves to notice that the non-trivial dependence
on $N_{APE}$ has to be carefully treated.
\subsection{Numerical results at $eB \neq 0$}\label{sec-2-1}
\begin{figure}[t] 
  \centering
 \includegraphics[width=0.45\textwidth]{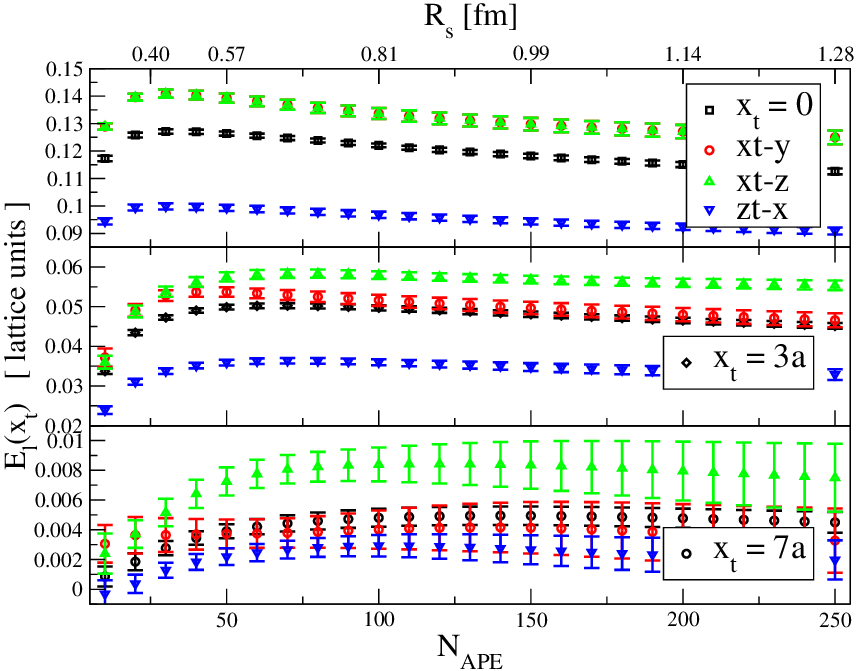}
  \caption{Dependence of the color-electric field strength on $N_{APE}$ (and on $R_s$) at $eB=0$ and at $eB\simeq\ 2{\rm GeV}^2$. Data refers to a $7\times 7$ Wilson Loop, \ie to an interquark separation of about $0.7$ fm.}
  \label{tube_B64_vs_nape_vs_Rs}
\vspace{-0.6cm}
\end{figure}

\begin{figure}[b] 
\vspace{-0.4cm}
  \centering
 \includegraphics[width=0.325\textwidth]{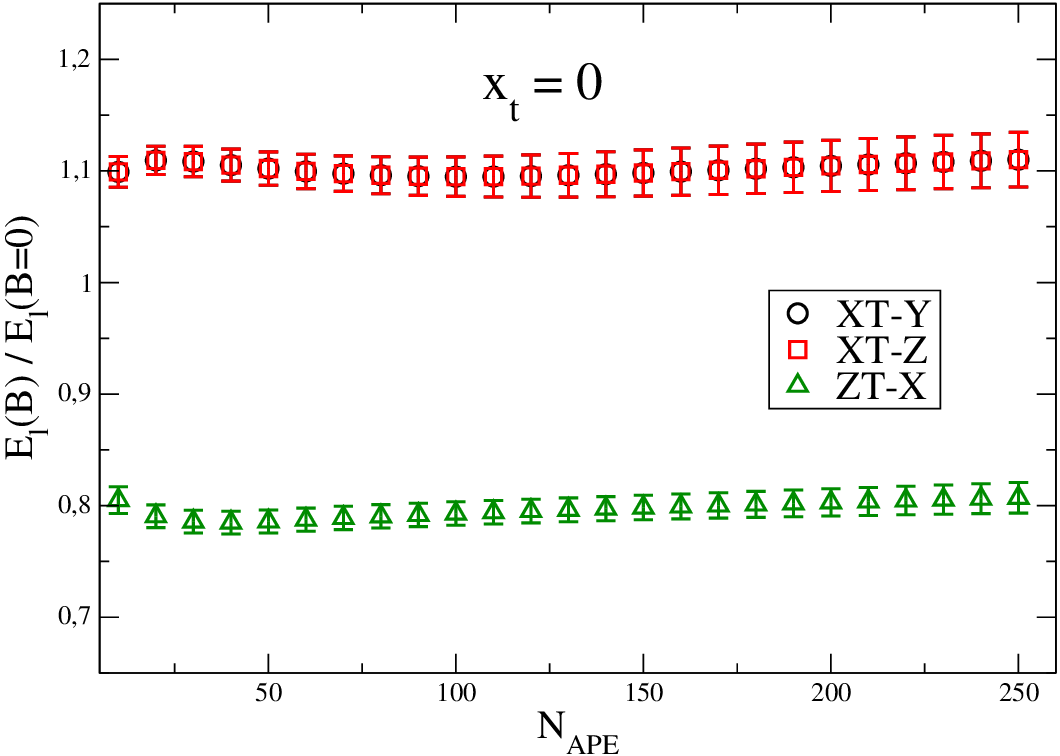}
 \includegraphics[width=0.325\textwidth]{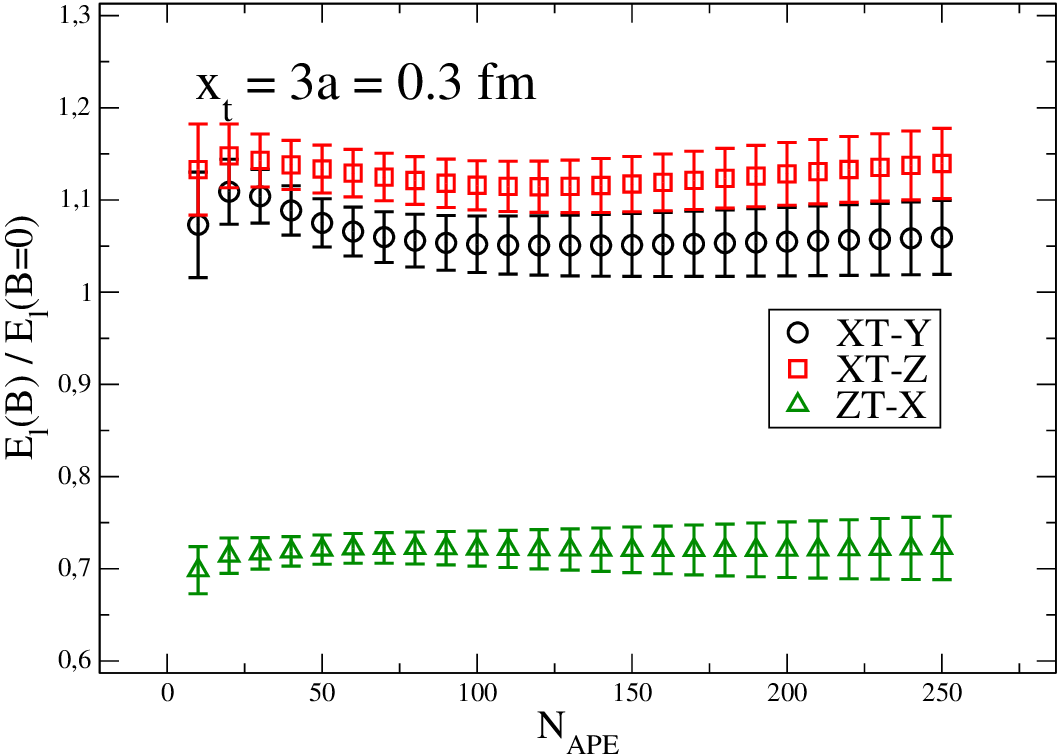}
 \includegraphics[width=0.325\textwidth]{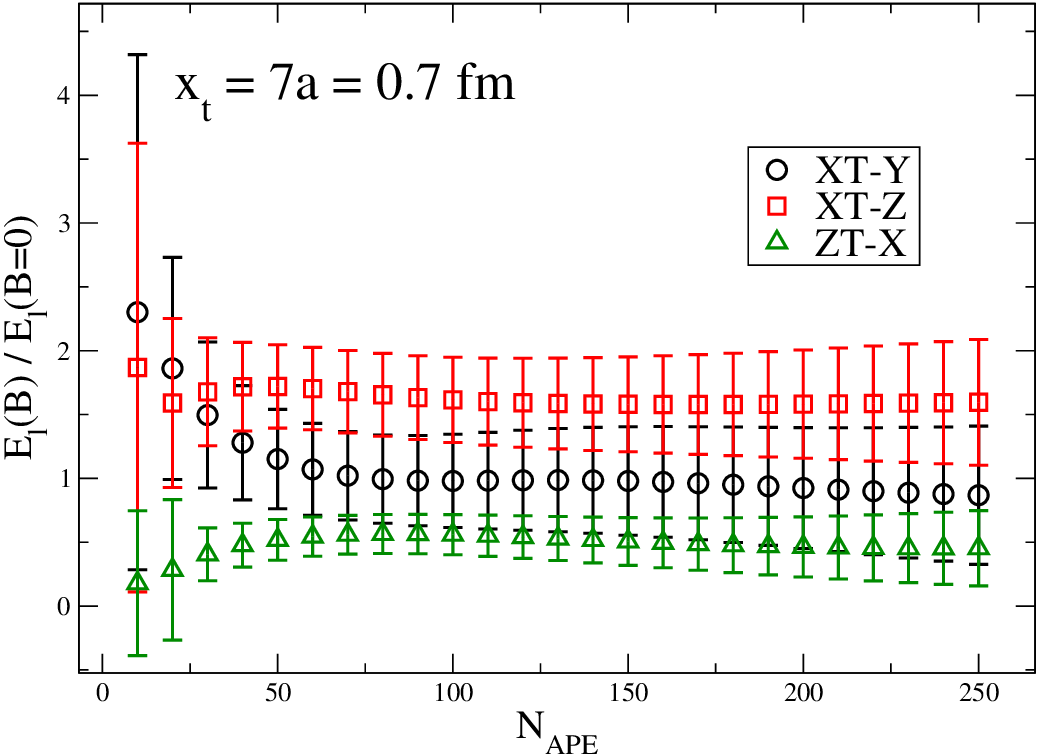}
  \caption{Ratios of the color-electric field strength at $eB=0$ and at $eB\simeq\ 2{\rm GeV}^2$, defined in Eq.~(\ref{ratios}),
at various transverse distances: {\it left} $x_t = 0$, {\it middle} $x_t = 3a$, {\it right} $x_t = 7a$. }
  \label{tube_vs_nape_ratios}
\end{figure}

\begin{figure}[h] 
  \centering
 \includegraphics[width=0.4\textwidth]{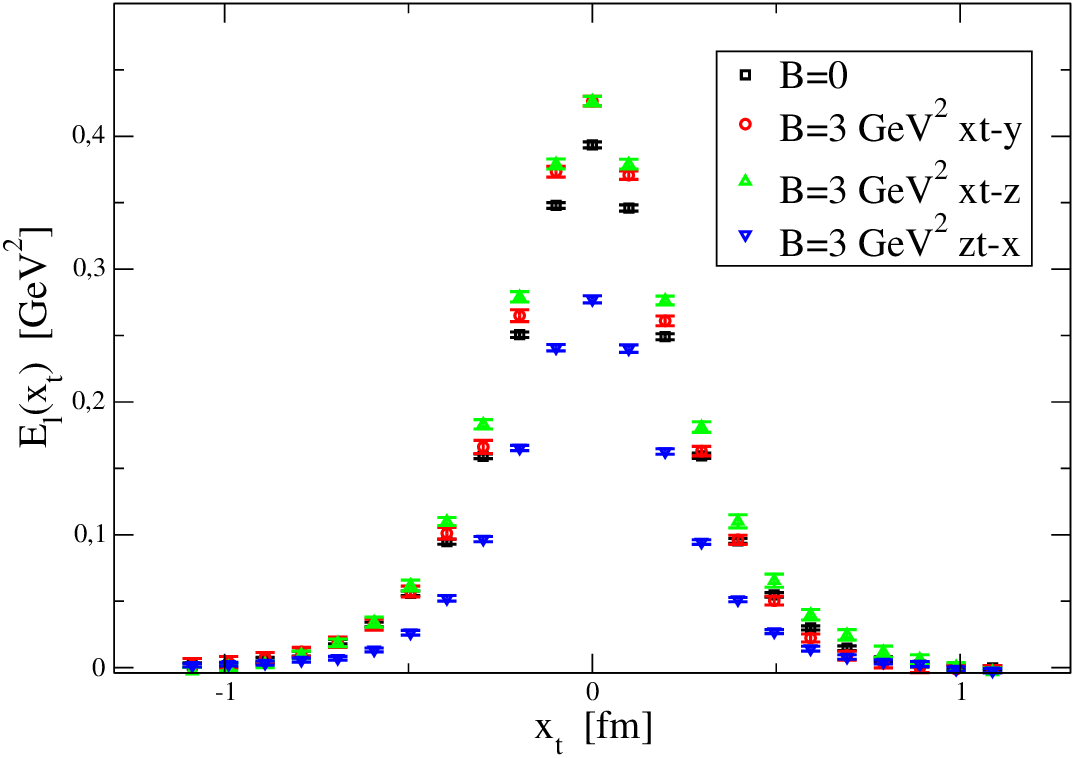}
 \includegraphics[width=0.4\textwidth]{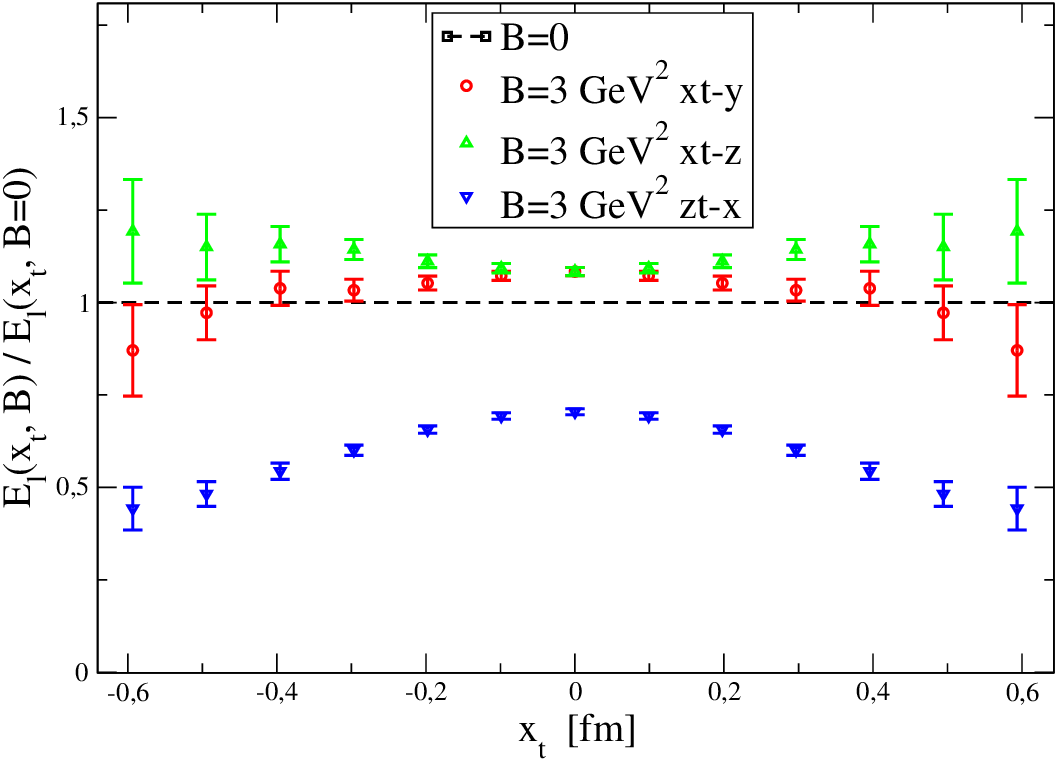}
  \caption{Flux tube profiles $E_{\ell}(x_t)$ at $eB=0$ and at $eB=3\ {\rm GeV}^2$. Black squares: $eB=0$. Red circles (green up triangles): $Q\overline{Q}$ separated along a direction orthogonal \wrt the external field, $x_t$ taken along a direction perpendicular (parallel) \wrt the field. Blue down triangles: $Q\overline{Q}$ separated along the  direction of the external field. {\it Left}: flux tube shape. {\it Right}: ratio of flux tube shapes, see Eq.~(\ref{ratios}).}
  \label{profile_B96}
\end{figure}

\vspace{-0.2cm}
In Fig.~\ref{tube_B64_vs_nape_vs_Rs} we compare the dependence on $N_{APE}$ (or on $R_s$)
of the field strength at $eB=0$ and at $eB \simeq 2\ {\rm GeV}^2$, as before for $x_t = 0,\ 3a,\ 7a$.
The results for three classes of direction combinations introduced in Sec.~\ref{sec-1} differs significantly:
the flux tube is modified in an anisotropic way.
In particular, as we separate the quarks along $X$ or $Y$ we observe an increase
of $F$ with respect to the $eB=0$ case, while it decreases as we separate the pair along $Z$,
the direction of the external magnetic field.
The problem of choosing a recipe to fix the number of smearing levels
still holds, indeed also data at non-zero external field display a
non-trivial dependence on $N_{APE}$.
Anyhow, such a dependence is quite similar as we go from $eB=0$ to $eB\neq 0$, while
staying at fixed $x_t$; the only difference between the two cases seems to be just a multiplicative factor.
To make this more quantitative, in Fig.~\ref{tube_vs_nape_ratios} we plot again the same data of
Fig.~\ref{tube_B64_vs_nape_vs_Rs}, but after having computed the ratios
\beq
\label{ratios}
r_{XT-Y} = \frac{E_{xt}(x_t\hat y,B)}{E_{xt}(x_t\hat y,B=0)}\ ,\ \ \ 
r_{XT-Z} = \frac{E_{xt}(x_t\hat z,B)}{E_{xt}(x_t\hat z,B=0)}\ ,\ \ \ {\rm and}\ \ \ 
r_{ZT-X} = \frac{E_{zt}(x_t\hat x,B)}{E_{zt}(x_t\hat x,B=0)}.
\eeq
All the curves are rather flat: the systematics introduced by the particular choice for $N_{APE}$
is almost washed away by the procedure of computing the ratios of Eq.~(\ref{ratios}).
In the left panel of Fig.~\ref{profile_B96}, we plot the profile of the flux tube both 
at $eB=0$ and all the 3 classes at $eB\simeq 3 {\rm GeV}^2$ when the color sources are
separated by $0.7$ fm.
Such data are not free from the systematics of the choice of the smearing level;
anyhow, we can appreciate how the field strength of the flux tube is more
intense if we separate the quarks along directions which are perpendicular
with respect to the magnetic field and, conversely, it is less intense
if we separate them along the direction of the field.
This anisotropy pattern is very similar and hence, likely, deeply related to that we observed
for the static $Q\overline{Q}$ potential in the presence of 
an external magnetic field~\cite{Bonati:2016kxj}, and specifically to the string tension anisotropy pattern.
In the right panel of Fig.~\ref{profile_B96}, we plot the ratio defined in Eq.~(\ref{ratios})
as a function of $x_t$. Given the independence of the ratios on $N_{APE}$ the plotted data
(which corresponds to $N_{APE}=80$) represents the value of the ratios at almost arbitrary $N_{APE}$.
This plot allows also to appreciate that the external field deforms the shape
of the flux tube, by shrinking or dilating its width.
Indeed, in the case of the $ZT-X$ field shape we observe that the ratio
is smaller than one at $x_t=0$ and gets smaller and smaller as $x_t$ increases:
the width is reduced by the magnetic field.
On the other hand, in the $XT-Y$ ($XT-Z$) case the ratio is larger than 1
in at $x_t=0$ and increases (decreases) as $x_t$ increases.
This means that, as we separate the quarks along a direction that is perpendicular
with respect to the magnetic field, the flux tube has no more a cylindrical symmetry:
it is shrinked along the direction of the field and dilated along the other direction (perpendicular to the field).
This loss of cylindircal symmetry is not particularly strong and
it becomes observable only at rather large external magnetic fields ($eB \sim 1\ {\rm GeV}^2$).\\
We fitted our data for the flux tube shape at $eB=0$ and at $eB\neq 0$ with the function
\beq
\label{clemfun}
E_{\ell}(x_t) = \frac{\phi\mu^2}{2\pi\alpha}\frac{K_0(\sqrt{\mu^2x_t^2+\alpha^2})}{K_1(\alpha)},
\eeq
which has been already adopted in the literature~\cite{Cea:2017bsa,Cea:2017ocq,Cea:2014uja,Cea:2012qw}, and that has been derived to describe the magnetic field profile of an Abrikosov vortex in an ordinary superconductor away from the London limit~\cite{Clem1975}.
In the context of QCD, such a function is intended to describe the shape
of the color-electric flux tube within the dual superconductor interpretation of the QCD vacuum.
The free parameters $\phi$, $\mu$ and $\alpha$ appearing in Eq.~(\ref{clemfun}) are, respectively,
the total flux, the inverse of the London penetration length and the ratio between the coherence length and a variational parameter.
\begin{figure}[b] 
  \centering
 \includegraphics[width=0.38\textwidth]{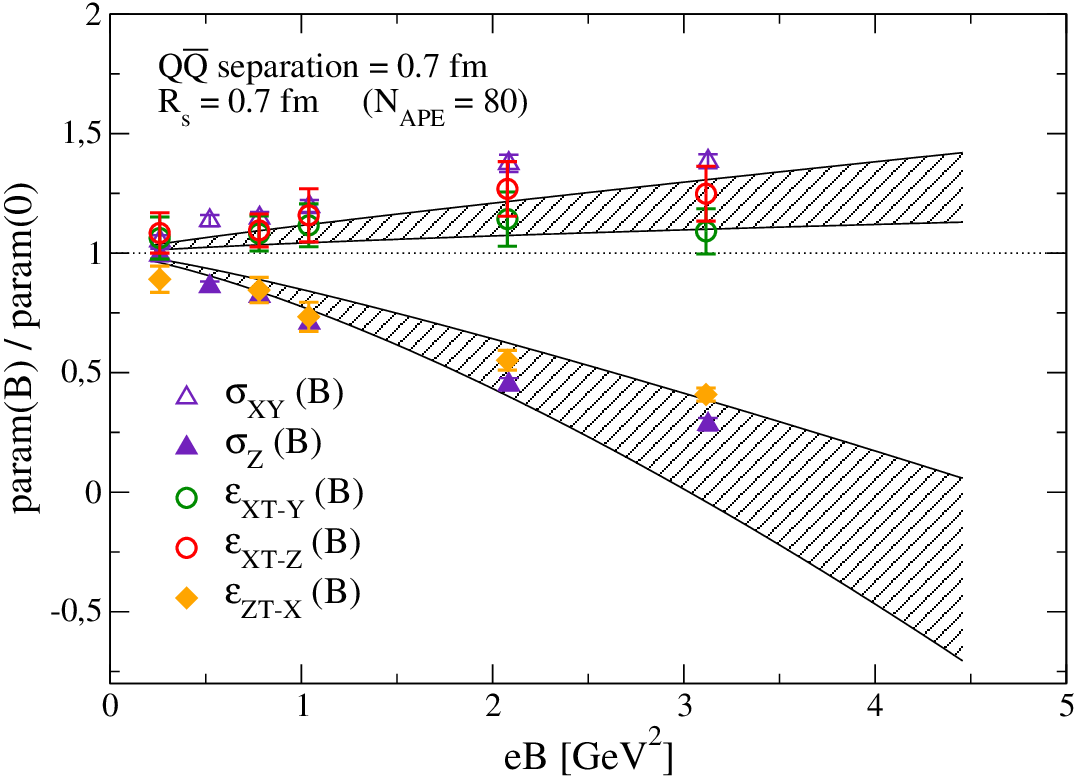}
  \caption{Full (empty) triangles: string tension ratios $\sigma_{Z}(B)/\sigma(B=0)$ ($\sigma_{XY}(B)/\sigma(B=0)$) as obtained in~\cite{Bonati:2016kxj} at the same lattice spacing we adopted in this work. Bands corresponds to the continuum limit of such ratios, always from~\cite{Bonati:2016kxj}.
We compare these results with that obtained in this paper.
Red (green) empty circles: flux tube energy density (see Eq.~(\ref{stringfun})) ratios $\epsilon_{XT-Y}(B)/\epsilon(B=0)$ ($\epsilon_{XT-Z}(B)/\epsilon(B=0)$). Orange full squares: $\epsilon_{ZT-X}(B)/\epsilon(B=0)$.}
  \label{sigwrat}
\end{figure}
Following~\cite{Cea:2017ocq}, we compute the value of the energy per unit length $\epsilon$ carried by the flux tube (\ie the string tension) by using the formula
\beq
\label{stringfun}
\epsilon = \int\!d^2\!x_t\ \frac{E^2_\ell(x_t)}{2} = \frac{\mu^2\phi^2}{8\pi}\left(1-\left(\frac{K_0(\alpha)}{K_1(\alpha)}\right)^2\right).
\eeq
To compare our results with the string tension ratio computed in~\cite{Bonati:2014ksa,Bonati:2016kxj}, we determined the ratio between the energy density $\epsilon$ at nonzero external magnetic field and that at zero field.
We report these ratios in the right panel of Fig.~\ref{sigwrat}, together with the string tension
ratios obtained from the static potential on the same lattice: this plot
shows that the two determinations are in agreement and supports the idea that there is a deep relation
between the two anisotropy patterns.
\section*{Acknowledgements}
We acknowledge PRACE for awarding us access to resource FERMI and MARCONI/A2 based in Italy at CINECA, under project Pra09-2400 - SISMAF and under INFN-CINECA agreement.
FN acknowledges financial support from the INFN SUMA project.
S.C. acknowledges support from the European Union’s Horizon 2020 research and innovation programme under the Marie Skłodowska-Curie grant agreement No. 642069.

\end{document}